\documentclass[aps,prl,twocolumn,showpacs,superscriptaddress,floatfix,nofootinbib]{revtex4}
\usepackage{amsfonts}

\usepackage{graphicx}
\usepackage{amsmath}
\usepackage{times}
\usepackage{amssymb}
\usepackage{color}
\usepackage[colorlinks,bookmarks=false,citecolor=blue,linkcolor=red,urlcolor=blue]{hyperref}

\begin{document}

\title{
Simultaneous dimerization and SU(4) symmetry breaking of 4-color fermions on the square lattice
}

\author{Philippe Corboz}
\affiliation{Theoretische Physik, ETH Zurich, 8093 Zurich, Switzerland}
\affiliation{Institut de th\'eorie des ph\'enom\`enes physiques, \'Ecole Polytechnique F\'ed\'erale de Lausanne, CH-1015 Lausanne, Switzerland}

\author{Andreas M. L\"auchli}
\affiliation{Max-Planck-Institut f\"{u}r Physik komplexer Systeme, N\"{o}thnitzer Stra{\ss}e 38, D-01187 Dresden, Germany}
\affiliation{Institut f\"ur Theoretische Physik, Universit\"at Innsbruck, A-6020 Innsbruck, Austria}

\author{Karlo Penc}
\affiliation{Research Institute for Solid State Physics and Optics, H-1525 Budapest, P.O. Box 49, Hungary}

\author{Matthias Troyer}
\affiliation{Theoretische Physik, ETH Zurich, 8093 Zurich, Switzerland}

\author{Fr\'ed\'eric Mila}
\affiliation{Institut de th\'eorie des ph\'enom\`enes physiques, \'Ecole Polytechnique F\'ed\'erale de Lausanne, CH-1015 Lausanne, Switzerland}

\date{\today}

\begin{abstract}
Using infinite projected entangled pair states (iPEPS), exact diagonalization, and flavor-wave theory, we show that the SU(4) Heisenberg model undergoes
a spontaneous dimerization on the square lattice, in  contrast to its SU(2) and SU(3) counterparts,
which develop N\'eel and three-sublattice stripe-like long-range order. Since the ground state
of a dimer is not a singlet
for SU(4) but a 6-dimensional irrep, this leaves the door open for further symmetry breaking. We provide
evidence that, unlike in SU(4) ladders, where dimers pair up to form singlet plaquettes,
here the SU(4) symmetry is additionally broken, leading to a gapless spectrum in spite of the broken translational symmetry.

\end{abstract}

%02.70.-c		:	Computational techniques; simulations
%71.10.Fd	:	Lattice fermion models (Hubbard model, etc.)
%03.67.-a		:	Quantum information
%67.85.-d,  ultracold Gases
%05.30.Fk, quantum statistical mechanics
%75.10.Jm Heisenberg model

\maketitle

The search for new quantum phases of matter, one of the main themes in the
field of electronic materials with strong correlations, has become an
important aspect of the physics of cold atoms too. A recent development concerns
the possibility of loading multi-color fermions in optical lattices, which
has reopened the investigation of the SU($N$) Heisenberg model that describes the
permutation of $N$-color objects on a lattice \cite{Wu03, Honerkamp04,Cazalilla09,Gorshkov10}. In condensed matter, these models require
fine tuning of parameters.
For instance, the SU(3) Heisenberg model is realized in spin-1 models with {\it equal}
bilinear and biquadratic interactions, while in standard antiferromagnetic materials the biquadratic
interaction is much smaller than the bilinear one. In the same spirit, the SU(4) model
can be seen as a symmetric version of the spin-orbital Kugel-Khomskii model~\cite{Kugel82},
but in actual materials the orbital-orbital interaction is not rotationally invariant \cite{materials}. By contrast, the simple
quantum permutation embodied by the SU($N$) Heisenberg model is a realistic starting
point to describe the Mott phase of $N$-color fermions at filling $1/N$ (one particle per site).

In this paper, motivated by the conflicting results published so far, we concentrate on the
case of 4-color fermions on the square lattice.
In the Mott insulating phase with one fermion per site and large on-site repulsion, the
relevant effective model is defined
by the Hamiltonian
\begin{equation}
\label{eq:H2}
\mathcal{H}= J \sum_{\langle i,j \rangle} P_{ij},
\end{equation}
where the transposition operator $P_{ij}$ exchanges two colors on neighboring sites, and $J>0$.

An investigation of the model based on a variational and mean-field study suggested a plaquette ground state \cite{Li98}, without a long-range color-order. Exact diagonalizations of 8 and 16 site clusters reported
the presence of low lying singlets~\cite{Bossche00}, and, inspired by the spontaneous singlet plaquette formation
in the SU(4) ladder~\cite{Bossche00b}, suggested that these low-lying singlets might live in the subspace of
singlet plaquette coverings, with possibly plaquette long-range order \cite{Hung11}. A different conclusion has been reached in Ref.~\onlinecite{Wang09}, where a gapless
spin liquid with nodal fermions has been shown to be the variational ground state for a class of projected fermionic wave functions.
The presence of a long-range color order with wave-vector ($2\pi/N$,$2\pi/N$)
for $N=2$ (N\'eel order) and $N=3$ (three-sublattice order with diagonal stripes~\cite{Toth10}) hints to the possibility of extending such an ordering to $N=4$. 
Lastly, a chiral spin liquid is predicted based on a particular large-N limit, but not below N=5 \cite{Hermele09}.
%Lastly, for $N\geq 5$ a chiral spin liquid is predicted based on a particular large-$N$ limit \cite{Hermele09}.

%%%%%%%%%%%%%%%%%%%%%%%%%%%%%%%%%%%%%%%
%%%%%%%%%%%%%%%%%%%%%%%%%%%%%%%%%%%%%%%
\begin{figure}[]
\begin{center}
\includegraphics[width=8.5cm]{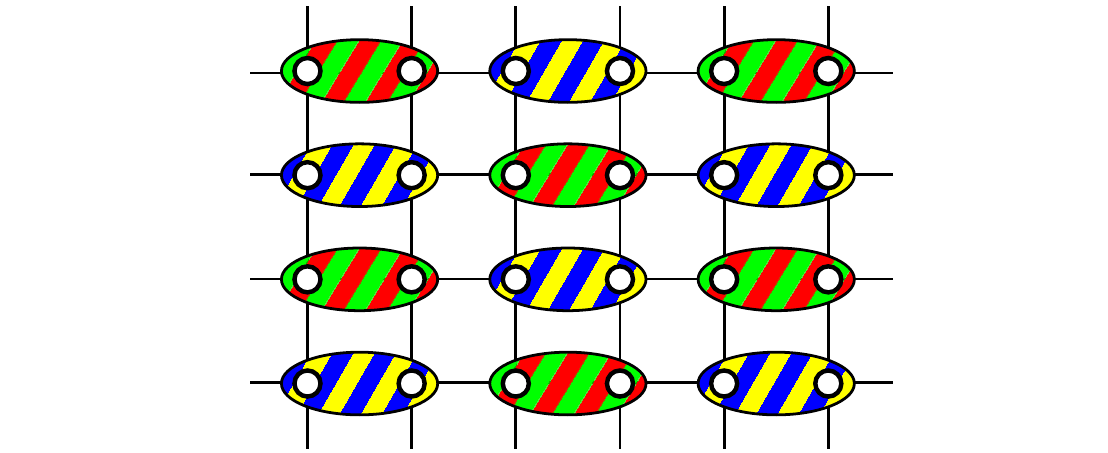}
\caption{(Color online) Sketch of dimer and color order realized
in the ground state of the SU(4) Heisenberg model on the square lattice.
}
\label{fig:finalpattern}
\end{center}
\end{figure}
%%%%%%%%%%%%%%%%%%%%%%%%%%%%%%%%%%%%%%%
%%%%%%%%%%%%%%%%%%%%%%%%%%%%%%%%%%%%%%%

In this Letter, we show that none of these possibilities is realized, and that in fact both
spatial and SU(4) symmetries are simultaneously broken: the ground state
is spontaneously dimerized, and on each strong dimer, two colors dominate and combine into a 6-dimensional irrep of SU(4). In the resulting pattern, the same color
does not appear on neighboring dimers, so that if two colors are dominant on one dimer, nearest-neighbor dimers
are dominated by the remaining two colors, resulting in the N\'eel-like arrangement sketched in Fig.~\ref{fig:finalpattern}.

\begin{figure}[t]
\begin{center}
\includegraphics[width=8cm]{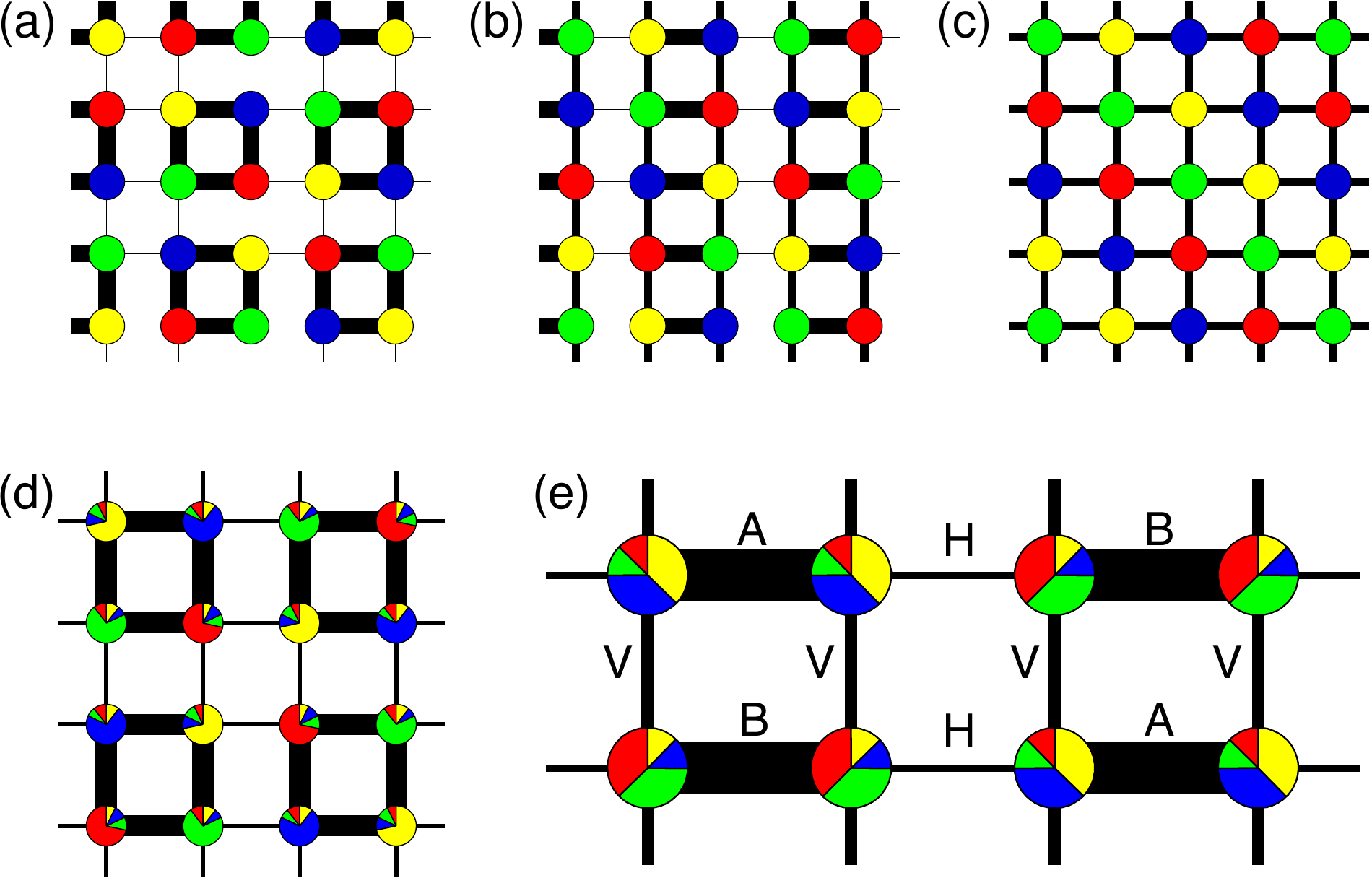}
\caption{(Color online) Sketch of some color and bond-energy configurations found by LFWT (a-c) and
iPEPS (d-e). In all plots, the thickness of the lines is proportional to the square of $\langle P_{ij} \rangle$
on the corresponding bond. The color content reflects that of the Hartree starting point for LFWT, and the actual
result for an appropriate, infinitesimal symmetry breaking field for iPEPS. (a) 'Plaquette' state selected
by quantum fluctuations within LFWT. Its energy per site $E_0/N =-3J/2$. (b) 'Dimerized' state with LFWT energy per site
$E_0/N = -1.293 J $. (c) Stripe-ordered state, with much higher LFWT energy per site $E_0/N = (4/\pi-2)J \simeq - 0.727 J$. (d) iPEPS results with $D=2$ and a unit cell $4\times 4$. At this order, the bond-energy pattern of the ground state is similar to that of the plaquette state selected by LFWT. (e) iPEPS results for  $D=12$ and a unit cell $4\times 2$. The bond-energy pattern of this state is similar to that of the dimerized state of LFWT. Two types of dimers A and B with energy much lower than vertical (V) and horizontal (H) dimers can be identified. They differ from each other by their local color densities.}
\label{fig:Hartree_configs}
\end{center}
\end{figure}

% variational description
%{\it Linear flavor-wave theory --}
The first hint that simple color ordering is probably not realized appears at the level of linear flavor-wave theory (LFWT),
the very method that confirmed the presence of stripe-order in the SU(3) case \cite{Toth10}. The starting point is a
Hartree approximation based on a site--factorized variational wave function. At this level, any state with different nearest-neighbor colors minimizes the energy, and, as in the case of SU(3), there is a macroscopic large number of such states. The next level of approximation consists in including quantum fluctuations within LFWT: The classical energy of each state is
corrected by the zero point energy of a
quadratic bosonic Hamiltonian which, on a nearest-neighbor bond with color $a$ on site $i$ and color $b$ on site $j$, is given by
$\mathcal{H}_\text{fw}= A^{\dagger}_{ij} A^{\phantom{\dagger}}_{ij} -1$, with $A^{\dagger}_{ij} = a^{\dagger}_{j} + b^{\phantom{\dagger}}_{i}$
%and $[A^{\dagger}_{ij},A^{\phantom{\dagger}}_{ij}]=0$.
where $a_j$ and $b_i$ are bosonic operators. Since there is an infinite
number of Hartree ground states, we have used the following strategy: compare simple, periodic states, and make a systematic exploration of all states on finite clusters. The resulting picture is summarized in Fig.~\ref{fig:Hartree_configs}(a-c). First of all, and most importantly, the state with stripe order of Fig.~\ref{fig:Hartree_configs}(c), which breaks SU(4) symmetry but leaves all nearest neighbor bonds equivalent, is clearly
{\it not} favored. In fact, many periodic states which break the spatial symmetry between bonds concomitantly with the SU(4) symmetry have much lower
energy, the lowest one being a
quadrumerized state in which strong bonds build 4-site plaquettes (Fig.~\ref{fig:Hartree_configs}(a)). Among possible states
we can also find a dimerized state (Fig.~\ref{fig:Hartree_configs}(b)) with energy slightly above that of the plaquette state but far below that of the stripe state.

This result, rather surprising in view of the SU(3) case~\cite{Toth10}, can be rationalized as follows: i) In a given ground state, the LFWT Hamiltonian is the sum of independent Hamiltonians that describe the motion of bosons on the connected clusters spanned by pairs of colors. For example, if a pair of neighboring sites with say colors $a$ and $b$ is such that all nearest neighbor sites have colors different from $a$ and $b$, it forms a 2-site connected cluster. For the state of Fig.~\ref{fig:Hartree_configs}(a), there are two types of clusters: 2-site clusters on strong bonds, and 4-site clusters (plaquettes) on weak bonds; ii) The zero point energy per bond tends to {\it increase} with the cluster size.
In particular, on a 2-site cluster, the ground state energy of the Hamiltonian is equal to $-1$, i.e. there is no zero-point contribution to the energy\cite{footnote1}, whereas larger clusters lead to finite frequencies, hence to strictly positive
contributions to the zero-point energy. In the stripe-ordered state of Fig.~\ref{fig:Hartree_configs}(c), all clusters are infinite, one dimensional zig-zag stripes, and no bond is able to minimize the energy. By contrast, half the bonds
have energy $-1$ in the state of Fig.~\ref{fig:Hartree_configs}(a) (thick bonds) and a quarter of them in
the dimer state of Fig.~\ref{fig:Hartree_configs}(b). The difference betweeen the present case and SU(3) comes from the fact that it is impossible to realize isolated two-site clusters with only three colors on the square lattice, hence to have local bonds with very low energy.

So flavor-wave theory gives strong indication that the ground state is not a simple color-ordered state, and
that some kind of additional lattice symmetry breaking takes place. One may be tempted to go one
step further, and to conclude that it predicts plaquette order. However, if the symmetry is broken and small
clusters appear, LFWT becomes inaccurate (for instance, it cannot restore the SU(4) symmetry in the ground states of a 4-site cluster), and one should rather turn to alternative methods.

% iPEPS description %
\paragraph{Infinite projected entangled-pair states (iPEPS) results ---}
Next, we present results obtained with iPEPS~\cite{PEPS}, generalized to arbitrary unit cells~\cite{Corboz11}, a variational approach based on a tensor network ansatz aimed at efficiently representing the ground state of two-dimensional lattice models in the thermodynamic limit. It consists of a network of rank-5 tensors, one for each lattice site, where each tensor is connected to its nearest neighbors by four bond indices with a certain bond dimension $D$, and the fifth index carries the local Hilbert space of a lattice site. The accuracy of the ansatz can be systematically controlled by the bond-dimension $D$. We use $N_T$ different tensors, arranged in a rectangular unit cell of size $L_x \times L_y=N_T$ which is periodically repeated in the lattice \cite{Corboz11}. The tensors are optimized by performing an imaginary time evolution, using the so-called simple update \cite{simpleupdate, comment_fullupdate}. For the computation of expectation values we use the corner-transfer matrix method \cite{CTM,Corboz11} to approximately contract the tensor network, where the accuracy of the contraction can be controlled by another parameter called boundary dimension~$\chi$. The simulation results presented in the following are extrapolated in $\chi$, where the extrapolation error is small compared to the symbol sizes.  To improve the efficiency for large bond dimensions  we use tensors with $\mathbb{Z}_q$ symmetry \cite{symmetrypapers} [a discrete subgroup of SU(4)].

\begin{figure}[]
\begin{center}
\includegraphics[width=8.5cm]{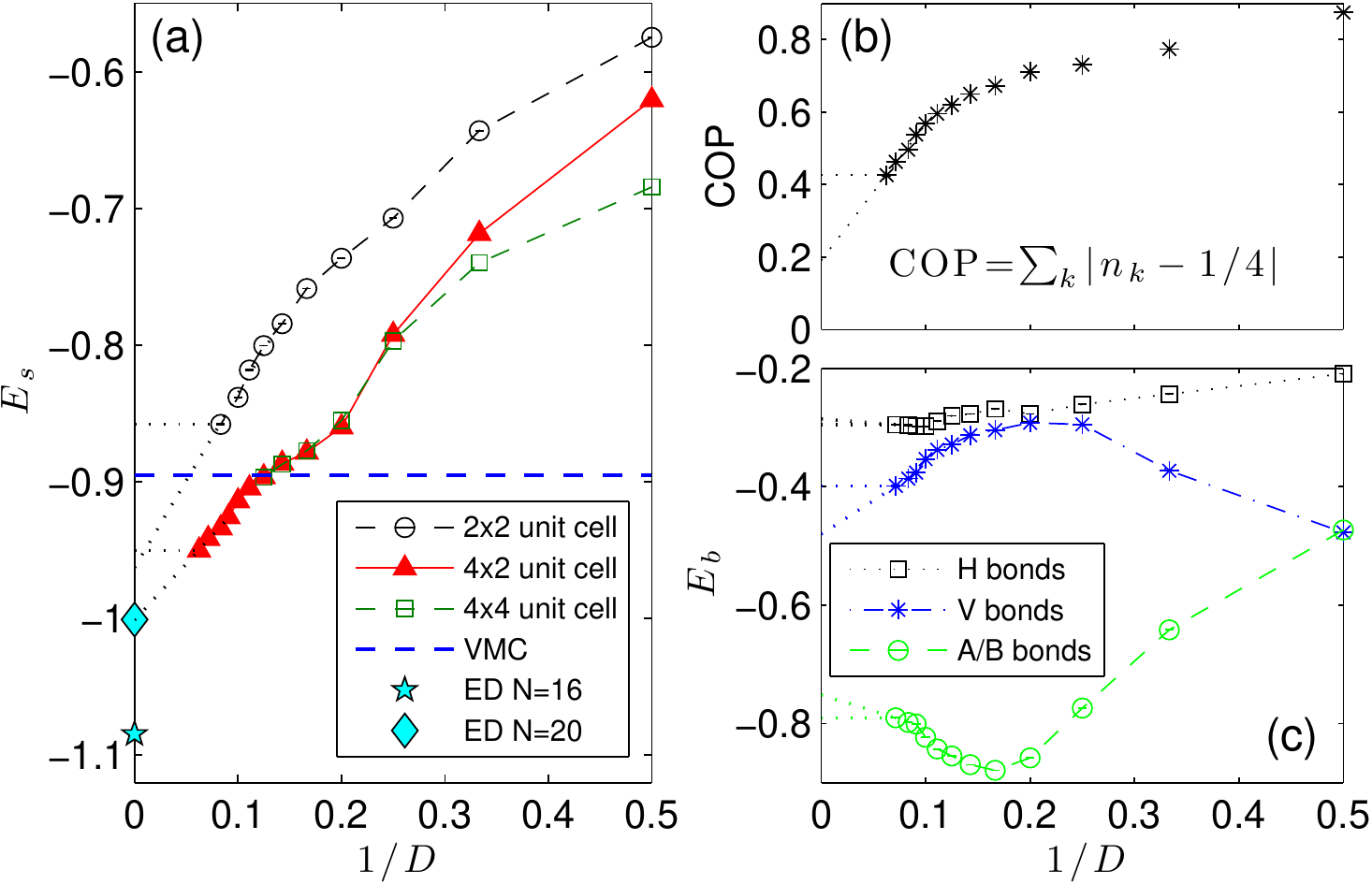}
\caption{(Color online) iPEPS results as a function of the inverse bond dimension $1/D$. Dotted lines are only a guide to the eye.
(a) Energy per site in units of $J$ for different unit cell sizes in the iPEPS, compared to the variational Monte Carlo (VMC) energy from Ref.~\cite{Wang09} (extrapolated to the thermodynamic limit), and exact diagonalization results. (b) Color order parameter (COP), which is zero if on each site the color density  $n_k$ of each color $k$ is the same, i.e. $n_k=0.25$. The data does not seem to extrapolate to zero in the limit $D\rightarrow \infty$ and we therefore expect the SU(4) symmetry to be spontaneously broken. (c) Bond energies of the bonds H, V, A and B marked in Fig.~\ref{fig:Hartree_configs}  obtained from the $4\times2$ unit cell for $D>4$ and from the $4\times 4$ unit cell for $D\le4$. The dimerization (strong A/B bonds) is seen to persist in the limit $D\rightarrow \infty$.
}
\label{fig:iPEPS}
\end{center}
\end{figure}

Figure~\ref{fig:iPEPS}(a) shows the iPEPS results for the energy per site as a function of $1/D$ for different unit cell sizes. The energy obtained with the $2\times 2$ unit cell is considerably higher than the one obtained with the $4\times2$ and $4\times4$ unit cells.  For $D\le4$ the lowest energy is obtained with the $4\times4$ unit cell, while for $D>4$ similar results are obtained with the $4\times2$ unit cell.
This indicates that the ground state breaks translational invariance in a way which is compatible with the $4\times2$ unit cell, but not with the $2\times2$ unit cell.
%Energy compared to VMC + remarks
For a bond dimension $D=8$ the iPEPS variational energy is comparable to the variational Monte Carlo (VMC) energy from Ref.~\cite{Wang09}, however, for larger $D$ we obtain energies which are considerably lower.
We note that the energy has not converged yet with bond dimension $D$, and thus further corrections to the ground state energy (and other quantities) can be expected for larger $D$.

 %%%%%%%%%%%%%%%%%%%%%%%%%%%%%%%%%%%%%%%
%%%%%%%%%%%%%%%%%%%%%%%%%%%%%%%%%%%%%%%
%%%%%%%%%%%%%%%%%%%%%%%%%%%%%%%%%%%%%%%
%%%%%%%%%%%%%%%%%%%%%%%%%%%%%%%%%%%%%%%

%local properties
We next turn our focus to local properties of the ground state. The bottom part of Fig.~\ref{fig:Hartree_configs} shows the local color densities on each site and the energy on each bond for $D=2$ (left) and $D=12$ (right). The evolution with $D$ of the energy of the various bonds is shown in Fig.~\ref{fig:iPEPS}(c).
For $D=2$, the bond pattern is similar to that of the linear FWT groundstate of Fig.~\ref{fig:Hartree_configs}(a). However, as soon as $D>2$, three types of bonds appear in the ground state, and the bonds form a columnar
dimerized pattern similar to the linear FWT of Fig.~\ref{fig:Hartree_configs}(b). Since increasing $D$ allows
more quantum fluctuations, we interpret this result as an indication that the dimerized pattern is ultimately
stabilized when enough quantum fluctuations are taken into account.

For SU(2) antiferromagnets, if spontaneous formation of dimers occurs, this is the end of the story: The ground state
degeneracy is given by the number of equivalent dimer coverings, and each ground state is adiabatically
connected to a product of singlets on the dimers. However, for SU(4), it takes at least 4 sites to make a singlet,
and on a dimer, the ground state is 6-fold degenerate. This 6-dimensional subspace corresponds to the 6-dimensional irrep of SU(4).
A convenient basis is provided by the states
$(\vert \alpha \beta \rangle - \vert \beta \alpha \rangle)/\sqrt{2}$ where the couple $(\alpha,\beta)$ is a pair
of different colors chosen out of the colors of the model and can take 6 values. 
These states are connected
by SU(4) rotations. So, even in the presence
of spontaneous dimerization, there is room for further SU(4) symmetry breaking. In our iPEPS results
for $D=12$, one can immediately recognize the formation of two types of dimers, marked with A and B in Fig.~\ref{fig:Hartree_configs}(e), which are arranged in a (columnar) checkerboard order. The energy of both A and B dimers is identical, however, the local color densities are different. In dimers A the first two colors [blue and yellow in Fig.~\ref{fig:Hartree_configs}(e)] are dominant, whereas in the B-dimers the density of the other two colors is larger. [Here we applied a small initial field to fix the "direction" of the SU(4) symmetry breaking in the space of the four colors. We note that the same results (up to SU(4) rotations) are obtained without an initial field].
This type of order is seen to persist also in the limit $D\rightarrow \infty$, as shown in Fig.~\ref{fig:iPEPS}(b).
We also verified that in the simulations done with a $4\times 4$ unit cell this pattern is repeated twice. Thus,
our iPEPS results suggest that both the SU(4) symmetry and translational symmetry are broken, resulting in a N\'eel-like order with
dimers alternating between pairs of colors.

%%%%%%%%%%%%%%%%%%%%%%%%%%%%%%%%%%%%%%%
%%%%%%%%%%%%%%%%%%%%%%%%%%%%%%%%%%%%%%%
\begin{figure*}[t]
\begin{center}
\includegraphics[width=\linewidth]{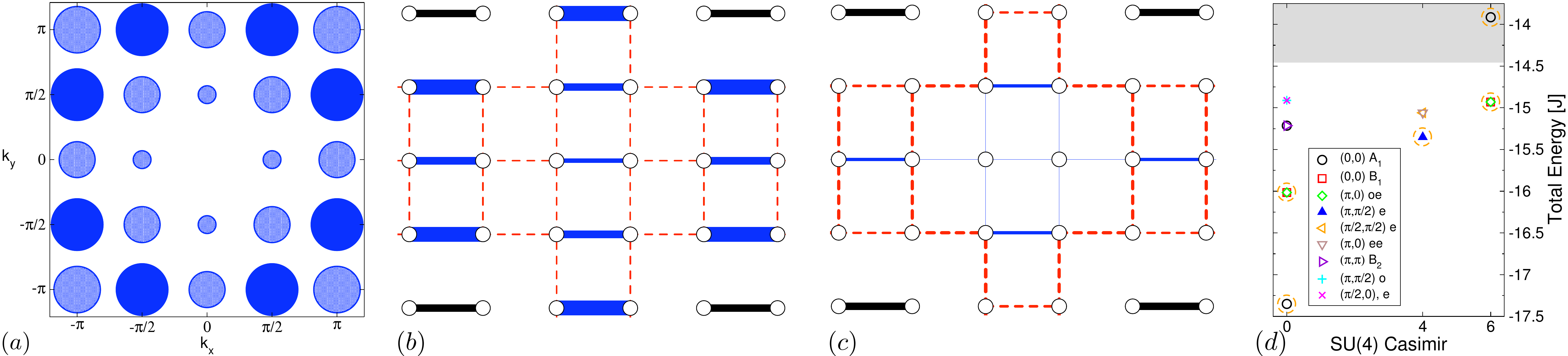}
\end{center}
\caption{(Color online)
(a) Color structure factor for the $N=16$ square sample in momentum space.
The maxima of the structure factor (filled circles) are located at momentum $(\pi,\pi/2)$ and
equivalent points.
(b) Connected bond energy correlations $\langle P_{ij} P_{kl}\rangle - \langle P_{ij}\rangle^2$ for $N=16$.
(c) $\langle \mathcal{P}_{(a,b)}(i,j)\mathcal{P}_{(a,b)}(k,l)\rangle$. In (b) and (c) the line width is proportional to the absolute value of
the correlations and blue, full (red, dashed) lines denote positive (negative) values. (d) Low energy spectrum of the $N=16$ sample
plotted as a function of the quadratic SU(4) Casimir operator $C$.
Expected levels in the tower of states of the SU(4) symmetry breaking state
sketched in Fig.~\ref{fig:finalpattern} are highlighted by dashed circles.
In the grey shaded region there are additional levels (not shown) which have not been
resolved as function of the Casimir.
}
\label{fig:corrs_ED}
\end{figure*}
%%%%%%%%%%%%%%%%%%%%%%%%%%%%%%%%%%%%%%%
%%%%%%%%%%%%%%%%%%%%%%%%%%%%%%%%%%%%%%%

%ED results
\paragraph{Exact diagonalization (ED) results ---}
We now compare these predictions with the results of ED of finite
clusters.
The energies per site $[E/(NJ)]$ in the ground state are $-1$ for $N=8$, $-1.0844$ for %25387$ for
$N=16$~\cite{Bossche00} and
%$-1.000753220$ for $N=20$~\cite{comment_N20}.
$-1.0008$ for $N=20$.
In Fig.~\ref{fig:corrs_ED} we present results for several correlation functions obtained
for the $N=16$ system. In Fig.~\ref{fig:corrs_ED}(a) Fourier-transformed color-density correlations are shown.
The diameters of the circles are proportional to the Fourier component. The
maxima are found at momentum $(\pi,\pi/2)$ and equivalent positions. This provides a possible
explanation as to why the $N=8$ and $N=20$ samples have higher energy per site than
the $N=16$ cluster, as their Brillouin zone does not contain these particular momenta and thus
frustrate the preferred order~\cite{comment_N20}. In Fig.~\ref{fig:corrs_ED}(b) we display the connected nearest-neighbor bond energy correlations
$\langle P_{ij} P_{kl}\rangle - \langle P_{ij}\rangle^2$. The correlations present a
clear striped signal and are in qualitative agreement with the bond energy modulation
pattern obtained using iPEPS, however the modulation is not very strong. Finally in Fig.~\ref{fig:corrs_ED}(c) we calculate
correlations between the projectors onto a given basis state of the $d=6$ representation
of SU(4) on a bond $\langle {\cal P}_{(a,b)}(i,j){\cal P}_{(a,b)}(k,l)\rangle$. This correlation
supposedly captures the N\'eel structure on top of a columnar dimerized background.
Indeed the correlations in (c) show the expected four positively correlated bonds at the
correct location confirming the qualitative picture put forward in Fig.~\ref{fig:finalpattern}.
We have also measured the spin chirality correlations advocated in Ref.~\cite{Hermele09}.
We indeed find correlations pointing towards uniform chirality on triangles, however the
correlations decay quickly over the size of the sample, implying the probable absence of
a chiral spin liquid phase on the SU(4) square lattice.

We conclude this ED section by presenting the low energy tower of state in Fig.~\ref{fig:corrs_ED}(d)
as a function of the quadratic SU(4) Casimir operator. In the case of either discrete or continuous
symmetry breaking the low energy spectrum is expected to exhibit characteristic energy levels which
become degenerate with the ground state in the thermodynamic limit and which enable the spontaneous
symmetry breaking. For the proposed state in Fig.~\ref{fig:finalpattern} we expect a tower of states
which is a fourfold superposition of the tower of state of a SU(4)$_6$ bipartite N\'eel state.
The later is given by the irrep content $C=0,4,6,10,12,\ldots$. Combining
with the spatial degeneracy associated with the four distinct dimer configurations, the even members of the
Casimir sequence are expected to belong to the spatial symmetry sectors $(0,0) A_1$, $(0,0) B_1$ and
the twofold degenerate sector $(\pi,0) \sigma_x=-1, \sigma_y=1$, while the odd members belong to the fourfold
degenerate momentum $(\pi,\pi/2)\ \sigma_x=1$ sector. When comparing this prediction (large circles) to the actual
low energy spectrum shown in Fig.~\ref{fig:corrs_ED}(d) one observes that the lowest levels for $C=0$ and $4$ are indeed the ones
expected for the proposed state in Fig.~\ref{fig:finalpattern}.

\paragraph{Conclusion ---}
The various approaches followed in this paper point to a rather coherent picture:  a spontaneous
dimerization appears in the ground state, and the remaining degrees of freedom, an irrep of dimension
6 on each dimer, develop N\'eel-like order. The model that describes the fluctuations around one of the
dimerized ground states should thus be a model with SU(4) symmetry with the 6-dimensional irrep at each
site of a square lattice and with anisotropic couplings - horizontal and vertical pairs of nearest-neighbor
dimers are inequivalent. 
In that respect, it is interesting to note that the isotropic version of the SU(4) Heisenberg model with the 6-dimensional irrep and only nearest-neighbor couplings has been studied by Quantum Monte Carlo~\cite{Assaad05} and
variational Monte Carlo \cite{Paramekanti07} with conflicting conclusions. While in the latter study long-range N\'eel order is found, the former predicts that the correlations decay algebraically, i.e. that there is no true long-range order.
Coming back to the evidence in favor of N\'eel order
in our case, the presence of short-range correlations is clear in view of the ED and iPEPS results. However,
correlations revealed by ED are not very strong, and the
prediction of long-range order relies mostly on the empirical $1/D$ scaling of the iPEPS order parameter of
Fig.~\ref{fig:iPEPS}. It would thus be very interesting to
challenge this prediction by deriving and studying the effective model relevant to the present case.

\acknowledgments We acknowledge the financial support of the Swiss National Fund and of MaNEP, and of the Hungarian OTKA Grant No. K73455. The simulations were performed on the Brutus cluster at ETH Zurich and on the PKS-AIMS cluster at the MPG RZ Garching.


\begin{thebibliography}{99}
\bibitem{materials} The $d$--electron system LiNiO$_2$ \cite{Kitaoka98} and the $f$--electron  system CeB$_6$ \cite{Shiina97} are believed to be quite close to an SU(4) symmetric model.
\bibitem{footnote1} On two sites, the Bogoliubov transformation is singular because
$[A^{\dagger}_{ij},A^{\phantom{\dagger}}_{ij}]=0$. However, the ground state energy can be shown
to be equal to -1, and the zero point energy to vanish because the frequencies of the two
underlying harmonic oscillators are equal to zero. A pair of zero frequency modes is actually
present on all finite clusters.
\bibitem{comment_fullupdate} We verified our results up to a bond-dimension $D=8$ with the so-called full-update (cf. \cite{Corboz10b}), which is optimal but computationally more expensive.
\bibitem{comment_N20}
The $N=8$ sample is a periodic square cell spanned by the two vectors $\mathbf{T}_1=(2,2)$ and $\mathbf{T}_2=(-2,2)$.
$N=16$ is spanned by $\mathbf{T}_1=(4,0)$ and $\mathbf{T}_2=(0,4)$ and
$N=20$ is spanned by $\mathbf{T}_1=(4,2)$ and $\mathbf{T}_2=(-2,4)$.
\end{thebibliography}
\end{document}